\documentclass{ws-procs9x6}

\begin{document}

\title{Monopoles, Duality, and String Theory}

\author{J. POLCHINSKI}

\address{Kavli Institute for Theoretical Physics, \\
University of California, \\ 
Santa Barbara, CA 93106-4030, USA\\ 
E-mail: joep@kitp.ucsb.edu}


\maketitle

\abstracts{
Dirac showed that the existence of magnetic monopoles would imply
quantization of electric charge.  I discuss the converse, and
propose two `principles of completeness' which I illustrate with
various examples.
Presented at the Dirac Centennial Symposium, Tallahassee, Dec. 6-7, 2002.}

\section{Theory}

It is a great honor to speak at this centennial of Paul Dirac. 
I also had the honor to speak at Pauli's centennial two years
ago and at Heisenberg's last year, and on each occasion it has been
interesting to go back and learn more about the work of these
men.  Of course the high point of their scientific lives
came rather early, with the discovery of quantum mechanics, and
the rest was anticlimax by comparison.  But from a modern
perspective the latter part is also fascinating, as they went on
to confront many problems that are still timely today.

All three thought very hard about the divergences of quantum field
theory, which was perhaps the central theoretical question of the
day.  On the other hand, Pauli and Heisenberg both looked for
unified theories, while Dirac regarded this as
premature.  It was his conviction that the immediate problem
facing theoretical physicists was to develop better mathematical
tools:\cite{dirac1}
\begin{quote}
The most powerful method of advance [is] to perfect and
generalize the mathematical formalism that forms the existing
basis of theoretical physics.
\end{quote}
Remarkably, not only did Dirac identify the necessary direction, he
followed it successfully and provided many key ideas that continue to play
a major role today:
\begin{itemize}
\item Magnetic monopoles.
\item Path integrals.
\item Light cone dynamics.
\item Membrane actions.
\item Conformal and de Sitter symmetries.
\item Constrained Hamiltonian dynamics.
\item Canonical formulation of gravity.
\end{itemize}
For an anticlimax that is a pretty good career.

Dirac comes across in many ways as the
first modern theoretical physicist.  Many of his statements
illustrate this, but the following strikes me as particularly
apt:\cite{varenna}
\begin{quote}
One must be prepared to follow up the consequences of theory, and
feel that one just has to accept the consequences no matter where
they lead.
\end{quote}
Dirac is often quoted on the importance of
mathematical beauty in one's equations; I did not choose one of these
quotations because beauty is so difficult to define.  He also made various
statements that one should not being distracted by experiment; I did not
choose one of these because they are inflammatory.

The reason that I find the chosen quotation so striking is that it is not
supposed to be possible to follow theory alone.  Without experimental
guidance, it is said, one will quickly become lost.  But of course today
in high energy theory we are to a large extent following theory where it
leads us, and we are rather confident that this is a correct and fruitful
path.  Why this approach can work is illustrated by Dirac's great
discovery:
\begin{equation}
\mbox{quantum mechanics} + \mbox{special relativity}\ \Rightarrow\
\mbox{antiparticles}\ .
\end{equation}
This was not a direct deduction (though in the framework of quantum field
theory, one can show that antiparticles are necessary for
causality.\cite{PeSc}).  Rather, when Dirac tried to find a consistent
framework that combined quantum theory and special relativity, he found it
very difficult --- so much so that when he did find one he had great
confidence in its inevitability, and was prepared to take its other
consequences seriously.

Essentially, with the discovery of quantum mechanics and special
relativity, and even more so with general relativity, theory has become
very rigid, so that it is difficult to extend or modify our existing
theories without making them inconsistent or otherwise unattractive. 
One reason for this is that relativity unifies space and time.  For
example, it seems almost inevitable that in quantum gravity space will be
modified or cut off at short distance.  One can imagine many sorts of
modification to the structure of space, but it is much harder to alter the
nature of time in a consistent way.

\section{The Necessity of Monopoles}

One of Dirac's remarkable discoveries was the connection between magnetic
monopoles and charge quantization.\cite{dirac1,dirac2}  Very early in the
history of quantum theory, he recognized the important connection between
geometry and quantum mechanics.  Dirac showed that in the presence of a
magnetic charge $g$, in order for the quantum mechanics of an electric
charge $e$ to be consistent one had to have
\begin{equation}
eg = 2\pi  n\ .
\end{equation}
Thus the existence of even a single magnetic charge forces every electric
charge to be a multiple of $2\pi/g$.  From the highly precise electric
charge quantization that is seen in nature, it is then tempting to infer
that magnetic monopoles exist, and indeed Dirac did so:\cite{dirac1}
\begin{quote}
One would be surprised if nature had made no use of it.
\end{quote}

I would like to discuss this from the point of view of the modern search
for a unified theory, and to offer two general principles of completeness:
\begin{enumerate}
\item 
In any theoretical framework that requires charge to be quantized,
there will exist magnetic monopoles.
\item 
In any fully unified theory, for every gauge field there will exist
electric and magnetic sources with the minimum relative Dirac quantum
$n=1$ (more precisely, the lattice of electric and magnetic charges is
maximal).
\end{enumerate}
Obviously neither of these is a theorem.  Rather, they are aesthetic
principles based on experience with a rather wide range of examples.  I
will give three examples of the first principle, and two of the second.

\subsection{Grand Unification}

The most well-known example of the first principle of completeness
is the 't Hooft-Polyakov monopole.\cite{tHP}  If the $U(1)$ of
electromagnetism is embedded in a semisimple group, for example in grand
unification
\begin{equation}
SU(3) \times SU(2) \times U(1) \subset SU(5)\ ,
\end{equation}
then electric charge is necessarily quantized, since it descends from the
quantized representations of the unified group.  Under precisely these
conditions, 't Hooft and Polyakov showed that magnetic monopoles will
exist as smooth but topologically nontrivial classical solutions.

Let me give a brief description of this idea.  Dirac showed that the
vector potential for a magnetic charge had a singularity along a string
extending from the charge.
\begin{figure}[th]
\epsfbox{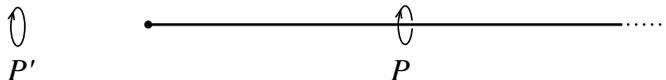}
\caption{A Dirac string extending from the monopole.}
\end{figure}
Figure~1 shows a Dirac string.  Let us
parallel transport a charged field $\psi$ on the infinitesimal path $P$
around it according to
\begin{equation}
d\psi = i e \vec A \cdot \vec dx \psi \ .
\end{equation}
Because of the string singularity, the field picks up a net phase $e
g$ in the process; this is unobservable precisely if $eg = 2\pi n$.  One
can think of the phase of $\psi$ as looping around the $U(1)$ group,
which is a circle, $n$ times.  Now, if we pull the loop off of the string
to the position $P'$, the field is nonsingular and the phase is constant.
Hence the net phase must drop rapidly from $2\pi n$ to zero as the loop
is pulled past the monopole, and this observable phase signifies a
singularity in the field. When $U(1)$ is embedded in a semisimple group,
however, the loop can become topologically trivial: in $SU(2)$ for
example, a rotation through
$4\pi$ can be smoothly deformed to a trivial path.  Thus there are
smooth field configurations, which at long distance look like Dirac
monopoles, with a net $U(1)$ magnetic charge.

As a postscript to the talk, I should note that this argument does not
require that one can obtain $eg=2\pi$, and indeed this depends on the field
content.  In the Georgi-Glashow model $U(1) \subset SU(2)$, if there are
only fields of integer isospin then the minimum quantum is $eg=4\pi$, while
if there are fields of half-integer isospin then $eg=2\pi$ is obtained. 
(This is related to the fact that some simple groups do have a finite set
of nontrivial closed loops.)  So in the sense of the second principle I
would have to say that this theory is not fully unified, precisely because
its matter content is not fixed but subject to arbitrary choice.

\subsection{Kaluza-Klein Theory}

If spacetime is five-dimensional, with the added dimension $x^4$
being periodic, then five-dimensional gravity gives rise to both gravity
and a Maxwell field in four dimensions.  This was perhaps the first
application of spontaneous symmetry breaking as a unifying concept --- the
laws of physics are invariant under Lorentz transformations in all five
spacetime dimensions, while the state we live in is invariant only under
the four-dimensional symmetry group.  The metric components $g_{\mu 4}$
become the Maxwell potential, and gauge invariance arises from
reparameterizations of $x^4$.  What the four-dimensional physicist
sees as electric charge is therefore momentum in the 4-direction, and it
is quantized because of the periodicity in this direction.

A Dirac monopole configuration would again have a string, which is now a
coordinate singularity as in Figure~1.  An infinitesimal loop in the
five-dimensional geometry, whose projection to four dimensions is the loop
$P$, makes one or more circuits of the $x^4$ coordinate.  An infinitesimal
loop away from the string, such as $P'$, does not loop the $x^4$
direction.  Thus there is again singular behavior as the loop is pulled
past the monopole.  However, Gross, Perry, and Sorkin\cite{kkmon} showed
that again there are smooth geometries that look like Dirac monopoles
outside of some core region.  The point is that if the radius of the $x^4$
direction shrinks to zero in an appropriate way, then at the origin there
is only a coordinate singularity and the loop can be smoothly slid off the
string.

\subsection{$U(1)$ Lattice Gauge Theory} 

The final example of the first principle is rather different from the
others, but very vividly illustrates the connection between charge
quantization and magnetic monopoles.  When one puts $U(1)$ gauge theory on
a spatial or spacetime lattice, the basic variables $A_l$ live on links,
directed pairs of adjacent points.  One can think of $A_l$ as the lattice
approximation to $\int \vec A \cdot d\vec x$, integrated from one site to
the next.
\begin{figure}[th]
\epsfbox{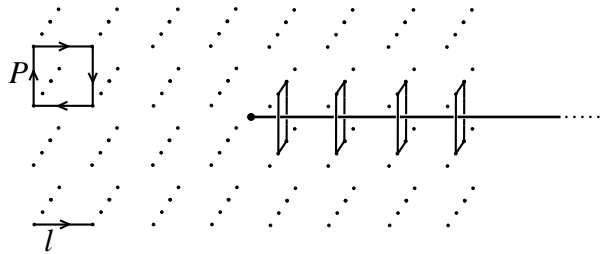}
\caption{Part of a spatial lattice.  A typical link $l$ and plaquette $P$
are illustrated.  A magnetic monopole and its Dirac string are shown,
hidden between the sites of the lattice.}
\end{figure}

There are two versions of $U(1)$ lattice gauge theory, according to
whether $A_l$ is a periodic variable or takes values on the whole real
line:
\begin{eqnarray}
\mbox{compact theory:}&&\quad A_l \cong A_l + 2\pi \ ,\nonumber\\
\mbox{noncompact theory:}&&\quad -\infty < A_l < \infty \ .
\end{eqnarray}
For example these theories have different actions.  The analog of the
field strength is the sum of the link variables around a plaquette,
\begin{equation}
F_P = \sum_{l \in P} A_l\ .
\end{equation}
In the noncompact case the simplest action would be the sum over
plaquettes of $F_P^2$, while in the compact case it would be the sum of $(1
- \cos F_P)$.  Alternately, one can describe the compact theory entirely in
terms of $U_l = e^{i A_l}$, and the action is constructed from 
$U_P = \prod_{l \in P} U_l$.

In the noncompact case charge is not quantized, one can transport any
charge $e$ from one site to the next with the phase $e^{i e A_l}$. 
However, in the compact case this is defined only for $e$ an integer, and
so charge is quantized (we are working here in units where the charge is
dimensionless and the minimum value is 1, but we can rescale $A_l$ to
other systems of units).  Precisely in keeping with the general principle,
one finds that there are magnetic monopoles in the compact case but not the
noncompact one.  The Dirac string is a line of plaquettes on all of which
$F_P \approx -1$ (or $-n$), while it is $\ll 1$ on all other plaquettes
except those very close to the monopole.
The line
ends at the monopole.  From the definition of $F_P$ it follows that $F_P$
summed over any closed surface is zero, so if we consider a large surface
surrounding the monopole the sum of all the small fluxes must be +1 (or
$+n$) to cancel the contribution of the string: this is the monopole flux.
Nothing is singular, everything is made finite by the lattice.  In the
compact theory the string costs no energy because $U_P$ is everywhere near
unity, while in the noncompact theory the string is visible to
noninteger charges and has an energy proportional to its length.

This example makes vividly clear the connection between charge
quantization and the existence of monopoles.  In fact, the compact theory
can be rewritten as the noncompact theory coupled to a magnetic monopole
field.  For some reviews of this subject see Ref.~4.

\subsection{The Kalb-Ramond Field}

Many supergravity theories have an antisymmetric tensor field
$B_{\mu\nu}$, with a generalized form of gauge invariance
\begin{equation}
\delta B_{\mu\nu} = \partial_\mu \lambda_\nu - \partial_\nu \lambda_\mu\ .
\end{equation}
If we consider such a theory in a Kaluza-Klein geometry,
the components
$B_{\mu 4}$ again become a four-dimensional Maxwell theory.  However,
unlike the Maxwell field from the metric, there are no states in
supergravity that are electrically charged under this gauge field --- there
is no way to minimally couple any field to $B_{\mu \nu}$ so as to give rise
to a minimal coupling to $B_{\mu 4}$.

This incompleteness is not an inconsistency, but it is somewhat puzzling
and unattractive that there is this asymmetry between the gauge fields
from the metric and those from $B_{\mu\nu}$.  In this respect string
theory completes supergravity.  The two-form $B_{\mu\nu} dx^\mu dx^\nu$
can be integrated over the world-sheet of the string, just as $A_\mu
dx^\mu$ can be integrated over the world-line of a particle, and just such
a coupling is present in string theory.  If a string wraps around the
periodic $x^4$ direction, the integral $dx^4$ produces a coupling to
$B_{\mu 4}$.  Thus the string winding states are electrically charged.

There should also be a corresponding magnetic source.  We can discuss this
in four dimensions, but it is clearer to start in the full ten-dimensional
theory.  In general a $p$-dimensional object (brane) couples to a
$(p+1)$-form potential, through the integral 
\begin{equation}
\int B_{\mu_1 \ldots \mu_{p+1}} dx^{\mu_1} \ldots dx^{\mu_{p+1}}
\end{equation}
over the world-history of the brane.  The curl of this potential gives a
$(p+2)$-form field strength.  Contracting this with the spacetime
$\epsilon$ tensor gives a dual $(D-p-2)$-form field strength, where $D$ is
the dimension of spacetime.  This corresponds to a $(D-p-3)$-form magnetic
potential, and so couples magnetically to a $p'$-dimensional object for
$p' = D-p-4$.  In other words, $p + p' = D-4$.  The familiar electric and
magnetic charges are simply $p = p' = 0$ in $D=4$, but the Dirac
quantization argument extends directly to all such
pairs.\cite{lattice,pdirac}

For the case at hand $D=10$ and $p=1$ and so $p'=5$: the magnetic object is
a five-brane.  This was found by Callan, Harvey, and Strominger\cite{CHS}
as a solitonic solution to the low energy field theory of string theory
--- it is the Neveu-Schwarz (NS) five-brane.  Curiously their first paper
found a five-brane charge of 8 Dirac units.  This is consistent,
but rather odd.  Shortly afterward the authors realized that they had used
inconsistent normalizations of $B_{\mu\nu}$, and found that the
magnetic charge is the minimum Dirac quantum.

\subsection{D-Branes}

The final example has great significance for me.  In string theory, in
addition to the NS-NS field $B_{\mu\nu}$ there are other `Ramond-Ramond'
form potentials.  The terminology refers to the fact that a closed string
state is the product of the states of its right- and left-moving
oscillations, so one can make a bosonic field out of two bosonic states
(these are called NS-NS) or out of two fermionic states (called R-R). 
Unlike $B_{\mu\nu}$ these forms do not couple to the fundamental string. 

However, string theory has extended objects of a distinctive type, the
D-branes.  These are like topological defects, with the notable property
that a (normally closed) string can end on them.  There were various
early discussions of strings with such fixed endpoints.  In particular it
was argued that these objects had to be included in string theory because
one encountered them if one took the normal open string theory and
followed it to small radius ($T$-duality).  For some time afterward these
objects were still regarded as curiosities, but in the wake of string
duality, in which R-R charges are necessary to fill out the multiplets,
it was realized\cite{RR} that the consistency of weak/strong duality and
$T$-duality required that the D-branes carry the R-R charge.  Further,
they exist with just the right dimensions to provide a full set of
electric and magnetic sources, even for nondynamical 9- and 10-form fields.

The calculation of $eg$ is interesting.  It is shown schematically in
Figure~3.  
\begin{figure}[th]
\epsfbox{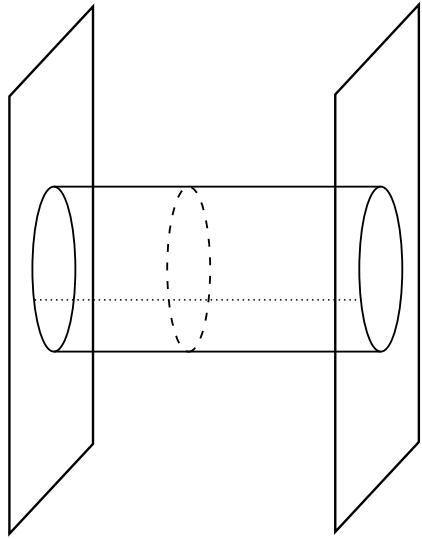}
\caption{Two D-branes (the vertical planes) and a cylindrical string
world-sheet with one boundary on each D-brane.  The world-sheet can be
regarded either as an open string with one end on each D-brane (the dotted
line) traveling in a vacuum loop, or a closed string (the dashed line)
emitted by one D-brane and absorbed by the other.}
\end{figure}
This process can be regarded as the emission and absorption of a
closed string, giving rise to a force between the two D-branes.  It can
equally be regarded as a vacuum loop of an open string, and this is by far
the most direct way to calculate it, by summing the zero point energies of
open string modes.  From the sum one reads off the potential, which gives
the coupling of the D-branes to the various forms.  Note that unlike the
calculations of the charges of the various other objects that have been
discussed there is nothing obviously topological about this, it is
essentially the calculation of a Casimir energy.  In particular there is
no reason that it should give anything like an integer for $eg$.  Indeed,
the first attempts at the calculation gave undesirable powers of $\pi$ and
$1/2$, but after a day of debugging the result is that the product of the
charge of a D$p$-brane and that of a D$(6-p)$-brane is exactly $2\pi$.

As a postscript, I should note that in any theory with gravity one can
make electric and magnetic sources trivially, by having the field lines
end on a black hole singularity (the R-R charged objects were first
described in this form\cite{bpb}).  In this description there is no direct
way to determine the actual spectrum of charges.  However in string theory
we can usually turn off the gravitational force by turning down the
coupling, so that the black hole goes over to something nonsingular, and
in this way we know the spectrum.

\section{Conclusions}

\subsection{The Existence of Monopoles}

By the end of his career Dirac became less certain about the existence of
monopoles.  He forgot his earlier dictum to ignore experiment!  But as I
have discussed, the existence of magnetic monopoles seems like one of the
safest bets that one can make about physics not yet seen.
It is very hard to predict when and if monopoles will be discovered.  If
their mass is at the grand unified scale as one expects, then they will be
beyond the reach of accelerators, while inflation has almost certainly
diluted any primordial monopoles beyond discovery.  It is curious to
contemplate this unfortunate situation, where theory predicts the
existence of an object (and its production, but in experiments that
can only be carried out in thought) and at the same time suggests that it
may never be seen.  But we must continue to hope that we will be lucky, or
unexpectedly clever, some day.

\subsection{Duality and Beauty}

Many of the electric and magnetic objects that I have discussed look
quite different from one another, but there is strong evidence that in
each case that there are dualities that interchange them.  For Dirac this
would have been a triviality.  He notes\cite{dirac2} that his theory of
pointlike electric and magnetic charges is invariant under the interchange
of the two objects, along with the interchange of electric and magnetic
potentials.  Dirac's theory is rather formal, since the magnetic
coupling is the inverse fine structure constant, but one can
regard the lattice as providing a precise definition of a cutoff theory,
and with appropriate choice of the action it is self-dual.\cite{lattice}

For the grand unified and Kaluza-Klein cases, however, any duality
must be quite nontrivial.  In these cases the electric charges are
pointlike quanta, while the magnetic charges are smooth classical
configurations.  To be precise, this is the picture at weak coupling.  Now
it would not be surprising that as the coupling is turned up the electric
objects begin to emit pairs and become big and fuzzy like the solitons. 
The great surprise (duality) is that when the coupling becomes very large
the magnetic objects become more and more pointlike and the theory can be
described in terms of their local fields.  It is a remarkable property of
the quantum theory that the degrees of freedom can, at least with some
assistance from supersymmetry, reorganize themselves in this way.  Indeed,
we do not fully understand the details of this, but the number of
independent consistency checks is enormous.

Even further, in string theory {\it all} of the electric and magnetic
objects that have been discussed here, with the (possible\cite{heller})
exception of the lattice examples, are related to one another by
dualities.  These examples involve widely different aspects of gauge field
geometry, spacetime geometry, `stringy' geometry, string perturbation
theory, and quantum and classical physics.  The existence of a single
structure that unifies such a broad range of physical and mathematical
ideas, and many others as well, is unexpected and remarkable.  Earlier I
declined to define beauty, but one can recognize it when one sees it, and
here it is.  This is one illustration of why the scientific path that Dirac
laid out has been such a fruitful one in recent times.

\section*{Acknowledgments}
This work was supported by
National Science Foundation grants PHY99-07949 and PHY00-98395.


\begin{thebibliography}{0}

\bibitem{dirac1} 
P.~A.~Dirac,
``Quantised Singularities In The Electromagnetic Field,''
Proc.\ Roy.\ Soc.\ Lond.\ A {\bf 133}, 60 (1931).

\bibitem{varenna}
P.~A.~Dirac, 1977 Varenna lecture, quoted by M.~Jacob 
in A.~Pais, M.~Jacob, D.~I.~Olive, and
M.~F.~Atiyah, ``Paul Dirac: The Man and his Work,'' Cambridge, UK: Univ.
Pr. (1998).

\bibitem{PeSc}
See for example Chapter 2 of M.~E.~Peskin and D.~V.~Schroeder,
``An Introduction To Quantum Field Theory,''
Reading, USA: Addison-Wesley (1995).

\bibitem{dirac2}
P.~A.~Dirac,
``The Theory Of Magnetic Poles,''
Phys.\ Rev.\  {\bf 74}, 817 (1948).

\bibitem{tHP} 
G.~'t Hooft,
``Magnetic Monopoles In Unified Gauge Theories,''
Nucl.\ Phys.\ B {\bf 79}, 276 (1974);\\
A.~M.~Polyakov,
``Particle Spectrum In Quantum Field Theory,''
JETP Lett.\  {\bf 20}, 194 (1974)
[Pisma Zh.\ Eksp.\ Teor.\ Fiz.\  {\bf 20}, 430 (1974)].

\bibitem{kkmon}
D.~J.~Gross and M.~J.~Perry,
``Magnetic Monopoles In Kaluza-Klein Theories,''
Nucl.\ Phys.\ B {\bf 226}, 29 (1983);\\
R.~d.~Sorkin,
``Kaluza-Klein Monopole,''
Phys.\ Rev.\ Lett.\  {\bf 51}, 87 (1983).

\bibitem{lattice}
M.~E.~Peskin,
Annals Phys.\  {\bf 113}, 122 (1978);\\
R.~Savit,
``Duality In Field Theory And Statistical Systems,''
Rev.\ Mod.\ Phys.\  {\bf 52}, 453 (1980).

\bibitem{pdirac}
P.~Orland,
``Instantons And Disorder In Antisymmetric Tensor Gauge Fields,''
Nucl.\ Phys.\ B {\bf 205}, 107 (1982);\\
R.~I.~Nepomechie,
``Magnetic Monopoles From Antisymmetric Tensor Gauge Fields,''
Phys.\ Rev.\ D {\bf 31}, 1921 (1985);\\
C.~Teitelboim,
``Monopoles Of Higher Rank,''
Phys.\ Lett.\ B {\bf 167}, 69 (1986).

\bibitem{CHS}
C.~G.~Callan, J.~A.~Harvey and A.~Strominger,
``Worldbrane Actions for String Solitons,''
Nucl.\ Phys.\ B {\bf 367}, 60 (1991);
``World Sheet Approach To Heterotic Instantons And Solitons,''
Nucl.\ Phys.\ B {\bf 359}, 611 (1991).

\bibitem{RR}
J.~Polchinski,
``Dirichlet-Branes and Ramond-Ramond Charges,''
Phys.\ Rev.\ Lett.\  {\bf 75}, 4724 (1995)
[arXiv:hep-th/9510017].

\bibitem{bpb}
G.~T.~Horowitz and A.~Strominger,
``Black Strings and P-Branes,''
Nucl.\ Phys.\ B {\bf 360}, 197 (1991).

\bibitem{heller}
S.~Hellerman,
``Lattice Gauge Theories have Gravitational Duals,''
arXiv:hep-th/0207226.



\end{thebibliography}
\end{document}